\documentclass[conference]{ieeeconf}
\IEEEoverridecommandlockouts
\usepackage{cite}
\usepackage{amsmath,amssymb,amsfonts}
\usepackage{algorithmic}
\usepackage{graphicx}
\usepackage{textcomp}
\usepackage{xcolor}
\usepackage{url}
\def\BibTeX{{\rm B\kern-.05em{\sc i\kern-.025em b}\kern-.08em
    T\kern-.1667em\lower.7ex\hbox{E}\kern-.125emX}}
\usepackage{comment}
\usepackage{amsmath}
\usepackage{amssymb}
\usepackage{mathtools}
\usepackage{booktabs}
\usepackage[dvipsnames]{xcolor}
\usepackage{dsfont}
\usepackage{pythonhighlight}
\usepackage{algorithm}
\usepackage{xspace}
\usepackage{tabularx}
\usepackage[acronym]{glossaries}
\usepackage{subcaption}
\usepackage{hyperref}
\hypersetup{
    colorlinks=true,
    urlcolor=green,      
    linkcolor=black,     
    citecolor=black      
}
\usepackage{balance}

\makeglossaries

\newacronym{vlm}{VLM}{vision language model}
\newacronym{llm}{LLM}{large language model}
\newacronym{av}{AV}{autonomous vehicle}
\newacronym{ai}{AI}{artificial intelligence}
\newacronym{DRA}{DRA}{Driving Record Analysis}
\newacronym{SG}{SG}{Scenario Generation}
\newacronym{RL}{RL}{reinforcement learning}

\definecolor{codegreen}{rgb}{0,0.6,0}
\definecolor{codegray}{rgb}{0.5,0.5,0.5}
\definecolor{codepurple}{rgb}{0.58,0,0.82}
\definecolor{backcolour}{rgb}{0.95,0.95,0.92}

\newcommand{\promptgeneric}[1]{{\color{blue} #1}}
\newcommand{\promptspecific}[1]{{\color{green} #1}}

\newcommand{\myred}[1]{{\color{red} #1}}
\newcommand{\mygreen}[1]{{\color{green} #1}}
\newcommand{\myblue}[1]{{\color{blue} #1}}

\newcolumntype{Y}{>{\centering\arraybackslash}X}

\title{\LARGE \bf From Driving Videos to Simulatable Scenarios}

\author{Alexandre Levy$^{1,2}$, Ernest Valveny Llobet $^{1,2}$, Antonio M. L\'opez$^{1,2}$ 
\thanks{$^{1}$Dept. Computer Science, Universitat Aut\`onoma de Barcelona (UAB)}%
\thanks{$^{2}$Computer Vision Center (CVC), UAB}
\thanks{\textbf{Accepted for publication at the IEEE International Conference on Intelligent Transportation Systems (ITSC), 2026.}}}

\newcommand{\ie}{{\em i.e.}}
\newcommand{\eg}{{\em e.g.}}

\newcommand{\etc}{{\em etc.}}

\usepackage{fancyhdr}   

\fancypagestyle{firstpage}{%
  \fancyhf{}%
  \fancyfoot[C]{\scriptsize\copyright~2026 IEEE. Personal use of this material is permitted. Permission from IEEE must be obtained for all other uses, in any current or future media, including reprinting/republishing this material for advertising or promotional purposes, creating new collective works, for resale or redistribution to servers or lists, or reuse of any copyrighted component of this work in other works.}%
}
\begin{document}

\maketitle
\thispagestyle{firstpage}
\pagestyle{empty}

\begin{abstract}
Autonomous vehicles (AVs) face driving scenarios ranging from routine traffic to rare events. To assess safety it is crucial to reproduce these scenarios in a controllable, repeatable, and scalable manner, with simulation playing a key role. This paper introduces D-V2S, a novel framework that automatically generates simulatable driving scenarios from driving videos. D-V2S operates in two stages: a Driving Record Analyzer (DRA) uses a vision language model (VLM) with our designed prompt to produce natural-language descriptions from input videos, capturing road layouts and dynamic traffic interactions; subsequently, a Scenario Generator (SG) uses a large language model (LLM) and our conditioning context to translate these descriptions into executable scenarios. Using simulations, we show that D-V2S generates scenarios where 90\% of the relevant semantic elements of the videos are present. We also provide qualitative results demonstrating D-V2S's capability to transform real-world driving videos into simulatable scenarios. Moreover, we provide both semantic and human driven ablative analyses of D-V2S's modules. In particular, we show how the VLM choice matters for DRA, and how our SG achieves a 75\% preference rate over other state-of-the-art methods.

\end{abstract}

\section{Introduction}

\Glspl{av} are rapidly becoming a reality and encounter diverse driving scenarios, from routine traffic to rare events. To assess safety, simulating these scenarios in a controllable and scalable manner is needed. Thus, generating simulatable targeted scenarios is core to develop the underlying \gls{ai} driving systems. 

To guide simulatable scenario creation, engineers analyze real-world driving 
videos from data-collecting vehicles, \gls{av} monitoring safety interventions, 
or prior simulations. The customized scenarios enable reliable simulation-in-the-loop incremental development of \glspl{av}, allowing evaluations of overall driving performance or specific \gls{ai} modules (perception, decision making, {\etc}).

However, the creation of scenarios involving particular events is a descriptive programming task \cite{asam_openscenario_v2, Fremont2019}. Thus, it requires engineers with specialized knowledge and becomes time consuming. Hence, we need a more automatic approach, which is the focus of this paper. We introduce a software framework, called D-V2S, that automatically transforms driving videos into simulation-executable driving scenarios. 

\begin{figure}[t]
    \centering
    \includegraphics[width=0.475\textwidth]{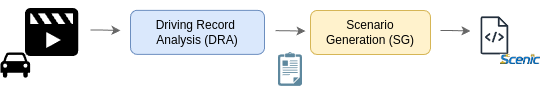}
    \caption{
    Two-stage D-V2S framework for automatic scenario generation: from a driving video, the \gls{DRA} module produces a natural-language description that is translated into a simulatable scenario by the \gls{SG} module.
    }
    \label{fig:pipeline}

\end{figure}

To design D-V2S, we started from practical considerations. On one hand, we decided to leverage \gls{vlm} technology for its high accessibility, automatic video interpretation, and continuous improvements. Moreover, we opt for using carefully designed prompts instead of VLM retraining, which can be costly in data and computation and requires coding. On the other hand, we decided to introduce a two-stage pipeline (Fig.~\ref{fig:pipeline}) with a \gls{DRA} and \gls{SG} to enable natural-language editable outputs. Specifically, the \gls{DRA} queries a \gls{vlm} with our prompt to generate a natural-language description of the driving video (road layouts, traffic participants, and dynamic interactions); then, inspired by prior works~\cite{LeGEND2024, ChatScene2024}, the \gls{SG} employs a context-conditioned \gls{llm} to transform this description into an executable scenario, allowing future natural-language modifications to add new desired elements anytime. Overall, if we need a simulatable scenario from a video, D-V2S (\gls{DRA}+\gls{SG}) performs the transformation fully automatically.

To conduct our research, we employ the SCENIC scripting language \cite{Fremont2019}, chosen for its open-source availability, high-level abstraction, support for randomization of scenario-parameter values (one code but multiple diverse simulations), and seamless integration with different simulators (MetaDrive, Webots, CARLA, {\etc}). Moreover, due to its widespread adoption and open-source nature, to run the scenarios generated by D-V2S, we use CARLA \cite{Dosovitskiy2017CARLA} as simulator. 

We assess the \gls{DRA} by comparing three representative \glspl{vlm}---LLaVA~\cite{LLAVA}, Qwen-VL~\cite{QwenVL}, and GPT-4o~\cite{GPT4}---using a prompt designed to elicit natural-language descriptions of input videos, including simulation-based failure cases of a driving \gls{ai} model. We evaluate them via objective semantic metrics and human judgment through elaborated surveys on the generated descriptions. GPT-4o clearly excels in accuracy and satisfaction, with statistical significance confirmed.

The assessment of the \gls{SG} compares it with two state-of-the-art methods, LCTGen~\cite{LCTGen} and ChatScene~\cite{ChatScene2024}, both for scenario generation from natural-language descriptions. Generated scenarios are rendered in bird's-eye view for comparison, using relevant descriptions of varying complexity, including some from the Crash Report Dataset~\cite{NHTSA2016CIREN}. Semantically, our method outperforms the others; its bird's-eye view outputs were preferred by judges in 75\% of cases via a designed survey, with statistical significance confirmed.

Finally, we assess D-V2S across testing videos presenting meaningful driving scenarios. Quantitatively, the generated scenarios preserve an average of 90\% of relevant semantic elements from the original videos. We also showcase D-V2S's qualitative results, including the use of real-world videos from a consumer-grade camera.

Hence, this paper makes the following contributions:
\begin{enumerate}
\item The use of a \gls{vlm} with a properly designed prompt to analyze driving video records and generate descriptions of the video content in natural language.
\item The use of a \gls{llm} with a proper context-conditioned prompt that permits to translate these descriptions into executable scenarios without manual programming.
\item The use of semantic metrics and statistically validated human judgments to quantitatively demonstrate the semantic fidelity and executability of scenarios generated by D-V2S. Moreover, we also assess the performance of \gls{DRA} and \gls{SG} modules individually.  
\end{enumerate} 

In the remainder of the paper, Section \ref{sc:relwor} situates D-V2S within the state of the art, Section \ref{sc:met} details D-V2S, Section \ref{sc:exp} describes our experiments and results, and Section \ref{sc:con} presents key messages and future directions.

\section{Related Works}
\label{sc:relwor}

Simulating realistic and diverse scenarios is a cornerstone of \gls{av} development. Simulators such as AirSim \cite{Shah2018}, CARLA \cite{Dosovitskiy2017CARLA}, LGSVL \cite{Rong2020}, and SUMO \cite{Lopez2018} support large-scale data collection and systematic evaluation of scenarios. However, they use default hand-crafted traffic rules that often fail to capture real-world traffic complexity. SceneGen \cite{SceneGen} introduced data-driven scenario generation, extended by TrafficGen \cite{TrafficGen}, with works \cite{Ghodsi2021, Ding2023} formalizing extraction and synthesis from large datasets for fine-grained control. 

Hand-crafted rules and statistical models lack flexibility to define complex intentions and semantic relations. In contrast, \glspl{llm} enabled reasoning over natural language to generate simulatable scenarios with richer complexity. LCTGen \cite{LCTGen} synthesizes traffic scenarios from natural language using a \gls{llm}-based interpreter for structured representations and a transformer-based generator for realistic multi-agent motions. ChatScene \cite{ChatScene2024} builds scenarios from unstructured language via a SCENIC snippet database: it decomposes instructions into sub-descriptions (behavior, geometry, spawn positions), embeds and retrieves code fragments, and assembles executable SCENIC scripts to produce physically consistent scenarios. ScenicNL \cite{ScenicNL2024} translates crash reports into SCENIC scripts using compositional prompting, constrained decoding, and compiler-in-the-loop feedback. More recently, Text2Scenario \cite{Text2Scenario2026} and Chat2Scenario \cite{Chat2Scenario2024} explore generating OpenSCENARIO \cite{asam_openscenario_v2} programs from language or structured inputs. Text2Scenario parses descriptions into logical forms assembled from predefined domain-specific language (DSL) fragments---reusable templates for elements like lane changes, accelerations, or yielding---producing deterministic executable scenarios. Chat2Scenario combines textual descriptions, criticality metrics, and trajectory datasets to retrieve and assemble scenarios under predefined maneuver taxonomies.

Other proposals translate videos into executable simulation code. Miao et al. \cite{dashcam2025} demonstrate converting dashcam collision clips to SCENIC scripts using a \gls{vlm}. LEADE \cite{LEADE2026} extracts abstract representations from traffic videos, producing instantiated scenario programs with fixed trajectories and timing. Road2Code \cite{Road2Code2025} uses multi-object tracking to recover precise vehicle trajectories, translated into SCENIC programs and evaluated via visual similarity metrics. However, small perception errors propagate directly into the generated scenario, and even minor modifications require rewriting trajectories or regenerating scripts.

Another emerging direction leverages multimodal \glspl{llm} for large-scale corner-case and environment generation. AutoScenario \cite{Autoscenario2026} synthesizes complete driving environments---road networks, agents, behaviors---from multimodal inputs, evaluating scenarios via textual similarity between input and output descriptions to explore diversity. While effective for diverse safety-critical cases, it prioritizes descriptive alignment and diversity over preserving specific event identity or causal structure, suiting corner-case exploration but not faithful replay or regression testing of observed events.

On the other hand, recent advances in \glspl{av} perception integrate \glspl{llm}/\glspl{vlm} with sensor data: CarLLaVA \cite{CarLLaVA2024} combines \glspl{vlm} and LLaMA \cite{LLAMA} for driving tasks; SimpleLLM4AD \cite{SimpleLLM4AD2024} provides driving descriptions; Xie et al. \cite{LLMreview2025} evaluate \glspl{vlm} across tasks, noting strengths and limits; CurricuVLM \cite{curricuvlm2025} generates curricula via reinforcement learning. However, these target perception/policy rather than scenario generation.

\textbf{In this paper}, we move toward a fully automated form of scenario generation by letting the input videos to drive the entire process. Unlike language-conditioned generators such as LCTGen, which synthesize multi-agent traffic motions from textual descriptions, or ChatScene and ScenicNL, which translate natural-language inputs into executable SCENIC scripts through code retrieval and compositional prompting, D-V2S does not require externally provided scenario descriptions. Similarly, in contrast to Text2Scenario and Chat2Scenario, which assemble OpenSCENARIO programs from predefined DSL fragments, maneuver taxonomies, or trajectory datasets, D-V2S does not commit early to fixed behavior templates or trajectory-level instantiations. Furthermore, in contrast to approaches such as AutoScenario, D-V2S explicitly preserves the identity and causal interaction structure of the observed event, {\ie}, focuses on semantic consistency. By operating in this semantic reconstruction regime, D-V2S bridges the gap between language-driven generation, trajectory-driven replay, and diversity-oriented synthesis. It enables interpretable, executable scenario generation suitable for debugging, benchmarking, and regression testing, thereby advancing the state of the art in real-to-simulation scenario generation.

\begin{figure}[h]
    \centering
    \includegraphics[width=0.475\textwidth]{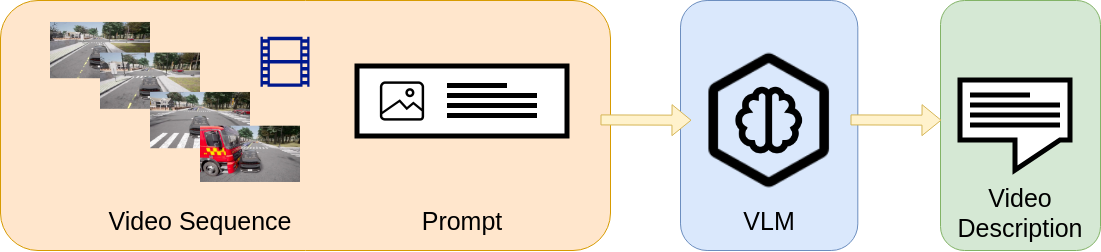}
    \caption{\gls{DRA}: it takes a video sequence and a prompt to output a natural-language description of the video content.}
    \label{fig:VLM_input}

\end{figure}

\begin{figure}[h]
\centering
\setlength{\fboxsep}{4pt}
\fbox{
\begin{minipage}{0.95\columnwidth}
\promptgeneric{This video shows footage from a rear-facing camera mounted on the ego vehicle.} 
\promptspecific{Why did the ego vehicle become blocked?}
\promptgeneric{\emph{Describe the behavior of the ego vehicle and the dynamics of the surrounding actors. 
Additionally, what type of road section is this---an intersection, a straight road, a highway, a T-intersection, or a three-way road?
Provide an answer in three sentences.}}
\end{minipage}
}
\caption{Text used in the \gls{DRA} module as \gls{vlm} prompt. The \promptgeneric{generic} component is highlighted in blue and the \promptspecific{video-specific component} in green, and the text emphasized in blue corresponds to the desired \promptgeneric{\emph{description of the output}}.}
\label{fig:prompt}

\end{figure}

\begin{figure}[t]
\centering
\setlength{\fboxsep}{4pt}
\fbox{
\begin{minipage}{0.95\columnwidth}
The \myred{ego vehicle} was \myblue{blocked by} a \myred{stationary digital signboard}  \myblue{placed on} the  \myred{right lane}, obstructing its path. The \myred{red car} \myblue{ahead of} the ego vehicle \mygreen{quickly maneuvered around} the signboard to continue on its way, while the ego vehicle \mygreen{remained stationary}. This scene takes place on a \myred{straight road} where the traffic dynamics cause a temporary halt in movement.
\end{minipage}
}
\caption{The \gls{vlm} answer contains \myred{traffic participants/elements} (red), their \mygreen{actions} (green), and \myblue{relationships} (blue).}
\label{fig:answer}

\end{figure}

\begin{figure}[t]
    \centering
    \includegraphics[width=0.475\textwidth]{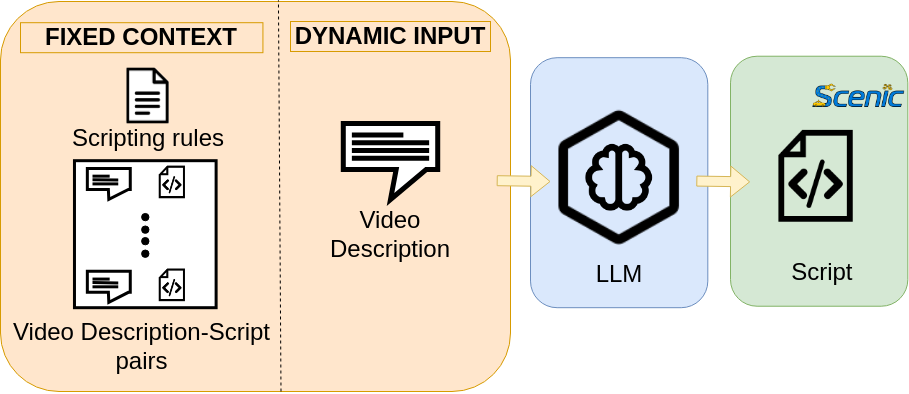}
    \caption{As input, \gls{SG} takes a fixed \emph{context} and a changing  (dynamic input) video description in natural language. The fixed context consists of the rules to script scenarios in SCENIC and pairs of a video description in natural language and a SCENIC script. The dynamic input describes the specific scenario to be generated and comes from the \gls{DRA}.}
    \label{fig:LLM_input}

\end{figure}

\section{Methodology}
\label{sc:met}

D-V2S converts videos from real or simulated driving into executable scenarios in the SCENIC scripting language, while also generating a natural-language description of the video content. It operates in two sequential stages (Fig.~\ref{fig:pipeline}): the Driving Record Analysis (DRA) and the Scenario Generation (SG). This section details DRA and SG.

\subsection{Driving Record Analysis (\gls{DRA})}
By using a \gls{vlm}, \gls{DRA} takes a driving video as input and produces a natural-language description of it as output. \Glspl{vlm} have input restrictions that require video sampling and frame resizing pre-processing steps. Moreover, \glspl{vlm} require a prompt (Fig.~\ref{fig:VLM_input}). Too open-ended prompts ({\eg}, \emph{Describe what happens in this video}) often produced generic or irrelevant outputs lacking spatial or causal details. Thus, we refined the prompt design by including contextual cues such as camera viewpoint and driving context ({\eg}, \emph{This video shows footage from a rear-facing camera mounted on the ego vehicle}), and explicit questions about the event of interest ({\eg}, \emph{Why did the ego vehicle become blocked?}). These adjustments improve the consistency and relevance of descriptions. Hence, our prompts consist of a general component common to all videos---informs the \gls{vlm} that the video depicts a traffic scenario and specifies relevant output information---and a video-specific one explaining its interest. Figure \ref{fig:prompt} shows an example. Then, the \gls{vlm} extracts details such as traffic participants/elements, their features, actions, relationships, {\etc} Figure \ref{fig:answer} shows a GPT-4o-based (VLM) description example.

\subsection{Scenario Generation (\gls{SG})}
\Gls{SG} takes the \gls{DRA} output---a natural-language traffic scenario description ({\eg}, Fig.~\ref{fig:answer})---and converts it into an executable SCENIC script using an \gls{llm}.

SCENIC is a high-level probabilistic programming language that declaratively specifies behaviors, spatial layouts, and environmental conditions for 
driving scenarios in simulators. A SCENIC script is organized as four self-contained sections that set: 
\begin{enumerate}
    \item The simulation map to use and global parameters such as the weather conditions or the ego-vehicle model. 
    \item The adversarial behavior that encodes how traffic participant nearby the ego-vehicle will act: lane change, sudden braking, {\etc}
    \item The road geometry under test: straight lane, T-junction, four-way intersection, {\etc}
    \item The spawn position for each traffic participant. 
\end{enumerate}

This is the structure that we aim to obtain as output of the LLM. However, as SCENIC is a relatively recent domain-specific language, there is a scarcity of publicly available scripts. Consequently, leveraging \glspl{llm} to directly generate such scripts is challenging, since these models have either not encountered this programming syntax during training or have been exposed to too few relevant examples to effectively learn its structure. In practice, when we prompted \glspl{llm} to generate SCENIC script, they frequently hallucinated non-existent language commands or omitted required components. Consequently, many of the generated scripts failed to compile, an issue also found in parallel studies on text-to-SCENIC generation \cite{ChatScene2024}.

To address the data scarcity without fine-tuning the \gls{llm}, we employ a lightweight prompt-engineering approach since the given \gls{llm} is provided with a fixed \emph{context} that emulates how a teacher start to explain a new programming language to a student. In particular, simple, diverse, and meaningful scripting examples are provided, {\ie}, pairs of one human-language description of the desired scenario and the SCENIC-compilable script that generates it. Moreover, the SCENIC scripting rules ({\ie}, the programming manual) are provided too as part of the contextual knowledge. The rules that we provide also include generic scripting pitfalls to avoid that we have collected during the development of the \gls{SG}. Note that many interfaces with \glspl{llm} allow to set this kind of common information as part of a working space so that only what changes (dynamic input) must be provided each time the \gls{llm} is prompted. Therefore, our approach follows the idea of in-context learning \cite{incontext} by guiding \glspl{llm} to generate SCENIC scripts that compile successfully, despite the language’s limited presence in the training data (Fig. \ref{fig:LLM_input}).

\section{Experiments}
\label{sc:exp}
We start by assessing the performance of \gls{DRA} and \gls{SG} individually: the former compares different \glspl{vlm} for driving scene interpretation; the latter measures how well SCENIC scenarios reproduce and generalize \gls{DRA}'s natural-language descriptions. Then, we assess the D-V2S pipeline (\gls{DRA} + \gls{SG}), quantifying original video semantics preserved in final executable scenarios. In addition, we show qualitative results.

\subsection{\glsdesc{DRA} (\gls{DRA})}

\subsubsection{\gls{vlm}}
Given the available \glspl{vlm}, we compared three competitive ones \cite{LLMreview2025}: LLaVA \cite{LLAVA} (LLaVA-Video-72B-Qwen2), Qwen-VL \cite{QwenVL} (Qwen2-VL-72B-Instruct), and GPT-4o \cite{GPT4} (gpt-4o via OpenAI API). LLaVA excels in open-source visual-linguistic reasoning, Qwen-VL in 
fine-grained visual perception, and GPT-4o represents the commercial frontier in multimodal reasoning.

\subsubsection{Multimodal Query}
To input video information to each \gls{vlm} within its per-request image limit, we sample frames at fixed frequency and time-limit while preserving temporal coverage. These frames are presented in temporal order with the textual prompt ({\eg}, Fig.~\ref{fig:prompt}), forming a multimodal query.

\subsubsection{Scenarios}
While developing AI models for AVs, we iterate through training and validation cycles. Failure-exposing scenarios are valuable for guiding subsequent improvements. In simulation, these arise during random or goal-oriented validations, making variations of failure-causing traffic events useful for retraining and regression testing.We follow this approach for validating the selected \glspl{vlm}.

\subsubsection{AI Model} 
To perform driving tasks, we selected CIL++, a fast, well-performing vision-based end-to-end driving model available on GitHub \cite{xiao2023_scaling}. CIL++ was trained on CARLA maps but, to our knowledge, not submitted to the CARLA Leaderboard challenge \cite{carla_leaderboard}.

\begin{figure}[t]
    \centering
    \includegraphics[width=0.475\textwidth]{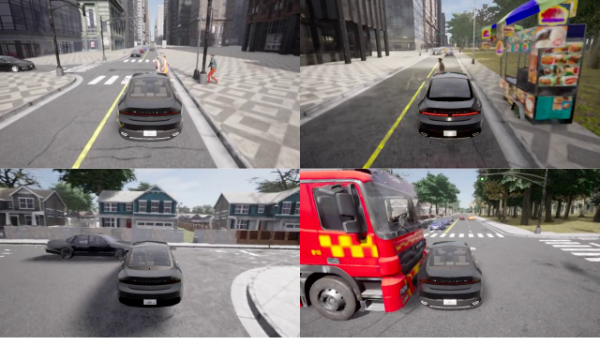}
     \caption{Examples of driving failures. Up row: running over pedestrians. Bottom row: vehicle crashes.}
    \label{fig:failure_images}

\end{figure}

\begin{table}[h]
\centering
\caption{Semantic Evaluation Metrics: Semantic Coverage Score (SCS), Hallucination Rate (HR), Semantic Preservation (SP), End-to-End Semantic Consistency (E2E-SC).}
\label{tab:semantic_metrics}
\begin{tabular}{l l p{4.5cm}}
\toprule
\textbf{Metric} & \textbf{Definition} & \textbf{Interpretation} \\
\midrule
SCS
& $ \frac{|GT \cap S|}{|GT|} $ 
& Recall of ground-truth video elements $(GT)$ in the generated description $(S)$. \\
\midrule
HR 
& $ 1 - \frac{|GT \cap S|}{|S|} $ 
& Fraction of generated elements $(S)$ not in the ground-truth $(GT)$. \\
\midrule
SP 
& $ \frac{|S_{de} \cap S_{sc}|}{|S_{de}|} $ 
& Recall of semantic elements in the natural-language description $(S_{de})$ preserved in the simulated scenario $(S_{sc})$. \\
\midrule
E2E-SC 
& $ \frac{|GT_{v} \cap S_{o}|}{|GT_{v}|} $ 
& Recall of ground-truth video elements $(GT_{v})$ preserved in the simulated output $(S_{o})$ after D-V2S execution. \\

\bottomrule
\end{tabular}

\end{table}

\begin{figure}
    \centering
    \includegraphics[width=0.475\textwidth]{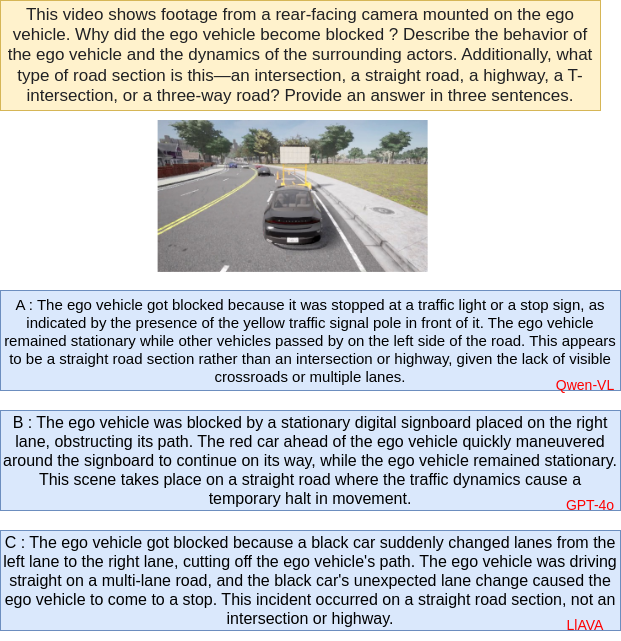}
    \caption{Example of different answers (blue boxes) produced by the three selected \glspl{vlm}, given a common prompt (yellow box) and recorded video (illustrative frame is shown here).}
    \label{fig:qos_failures}

\end{figure}

\subsubsection{Videos} 
Thus, we assessed CIL++ on the CARLA Leaderboard, recording videos of driving failures---specifically, seconds prior to each failure to capture preceding situations (Fig.~\ref{fig:failure_images}). These videos were recorded from a camera behind and slightly above the ego-vehicle (driven by CIL++).

\subsubsection{Experimental Setup and Evaluation Metrics} Each recorded video was processed with the three selected \glspl{vlm}, yielding three descriptions per scenario. For objective semantic extraction assessment, we annotated ground-truth semantic elements per video---covering traffic participants, road topology, and actions---from which we extracted predicted semantic sets from each VLM description to compute \emph{Semantic Coverage Score (SCS)} and \emph{Hallucination Rate (HR)} (Table~\ref{tab:semantic_metrics}). These metrics are deterministic and computed directly from annotated semantic sets; therefore, no hypothesis testing is required for their interpretation. Higher SCS and lower Hallucination Rate indicate better semantic grounding. This quantifies how completely and accurately each VLM captures scene structure. 

In addition to semantic evaluation, we conducted human evaluation with questions like those in Fig.~\ref{fig:qos_failures}. Highly experienced drivers assessed each scenario by selecting the best description of video content (\emph{Preference}) and rating each answer's adequacy on a 1--5 scale (\emph{Adequacy}). They were not informed of which \gls{vlm} generated each answer. To mitigate location bias, \gls{vlm} answer order was randomized per question. We prepared 62 questions for distinct scenarios, randomly split into two fixed surveys; they could complete one/both online (no time limit) and skip questions. In total, we collected 360 responses across surveys.

To assess statistical significance in human judgment we compute 95\% confidence intervals (CI) for true means, Chi-square tests for preference distribution differences, two-proportion $z$-tests for pairwise model comparisons, one-way ANOVA for adequacy score differences, Tukey post-hoc tests for specific significant pairs, and Cohen's $d$ for effect size (in standard deviations). These statistics are applied to both Preference and Adequacy answers.

\begin{table}
\centering 
\setlength{\tabcolsep}{4pt} 
\caption{\Gls{vlm} semantic description evaluation.} 
\label{tab:vlm_semantic} 
\begin{tabular}{@{}lcc@{}} 
\toprule 
\textbf{Model} & \textbf{SCS $\uparrow$} & \textbf{HR $\downarrow$}\\
\midrule 
LLaVA  & 0.70 $\pm$ 0.35          & 0.20 $\pm$ 0.23\\ 
GPT-4o & \textbf{0.91 $\pm$ 0.18} & \textbf{0.06 $\pm$ 0.17} \\ 
Qwen   & 0.69 $\pm$ 0.29          & 0.15 $\pm$ 0.22\\ 
\bottomrule 
\end{tabular} 

\end{table}

\begin{table}[h]
\centering
\setlength{\tabcolsep}{2.5pt}
\caption{\gls{vlm} evaluation by the surveyed.}
\label{tab:human_eval}

\begin{tabular}{lcccc}
\toprule
& \multicolumn{2}{c}{\textbf{Preference}} 
& \multicolumn{2}{c}{\textbf{Adequacy}} \\
\cmidrule(lr){2-3} \cmidrule(lr){4-5}
\textbf{Model} 
& Preference (\%) & 95\% CI (\%) 
& Mean $\pm$ St.Dev. & 95\% CI \\
\midrule
Qwen-VL & 19.17 & 15.43--23.55 & 2.53 $\pm$ 1.42 & [2.38, 2.68] \\
LLaVA   & 22.22 & 18.23--26.80 & 2.61 $\pm$ 1.35 & [2.47, 2.75] \\
GPT-4o  & \textbf{58.61} & 53.46--63.58 & \textbf{3.58 $\pm$ 1.26} & [3.45, 3.71] \\
\bottomrule
\end{tabular}

\vspace{0.25ex}
{\scriptsize
\textbf{\textit{Preference:}}\\
\textbf{\textit{Chi-square test:}}\\ 
$\chi^2(2, N=360)=104.017$, $p=2.589\times10^{-23}$, Cohen's $w=0.538$.\\
\textbf{\textit{z-test (p-value):}} \\
Qwen-VL vs LLaVA: $p=0.961$; \\
Qwen-VL vs GPT-4o: $p=7.98\times10^{-25}$; \\
LLaVA vs GPT-4o: $p=6.36\times10^{-21}$.

\textbf{\textit{Adequacy:}}\\
\textbf{\textit{One-way ANOVA:}} 
$F(2,1077)=67.8$, $p=1.8\times10^{-28}$, $\eta^2=0.112$.\\
\textbf{\textit{Post-hoc Tukey test (p-value) / Cohen's $d$:}} \\ 
Qwen-VL vs LLaVA: $p=0.73$, $d=-0.06$; \\
Qwen-VL vs GPT-4o: $p<0.001$, $d=-0.78$; \\
LLaVA vs GPT-4o: $p<0.001$, $d=-0.75$. \\
}
\end{table}

\begin{figure}[t]
    \centering
    \includegraphics[width=0.475\textwidth]{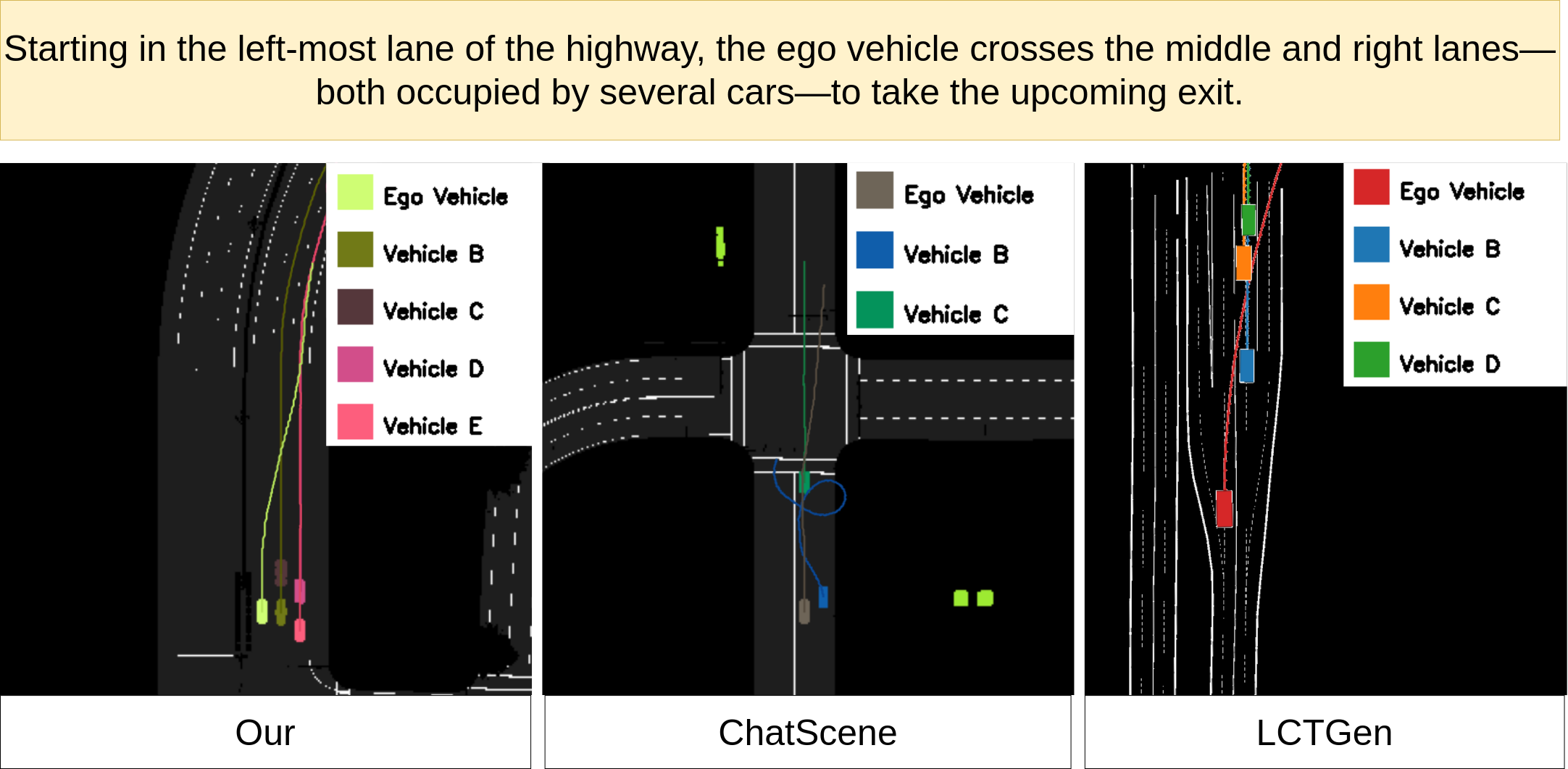}
    \caption{Given the description in the yellow box, we show the bird-eye-view of the generated scenarios from \gls{SG} (our method), ChatScene \cite{ChatScene2024} and LGTGen \cite{LCTGen}.}
    \label{fig:qos_generation}

\end{figure}

\subsubsection{Results and Discussion}
Table~\ref{tab:vlm_semantic} shows GPT-4o achieving the highest SCS (0.91), indicating nearly all elements from the annotated ground truth were detected, and with the lowest HR (0.06), indicating minimal mention to unsupported elements. Table~\ref{tab:human_eval} shows the preferences across the three \glspl{vlm}: GPT-4o was preferred in 58.6\% of cases (95\% CI: 53.46--63.58\%), Qwen-VL in 19.2\%, and LLaVA in 22.2\%. The CIs indicate where the true preference proportions likely lie; GPT-4o's substantial lead is statistically significant. A chi-square test confirmed deviation from equal preference ($\chi^2(2,N=360)=104.0$, $p<10^{-22}$, Cohen's $w=0.54$). Two-proportion $z$-tests showed GPT-4o significantly outperforming Qwen-VL ($p<10^{-24}$) and LLaVA ($p<10^{-20}$). Table~\ref{tab:human_eval} also reports adequacy scores: GPT-4o achieved the highest mean ($3.58 \pm 1.26$), while Qwen-VL and LLaVA received lower scores. One-way ANOVA confirmed significant differences among models ($F(2,1077)=67.8$, $p=1.8\times10^{-28}$, $\eta^2=0.112$). Tukey post-hoc tests showed GPT-4o rated significantly higher than both others ($p<0.001$). Cohen's $d$ also indicated large differences of GPT-4o with the others ($|d|\approx0.75$).

Considering these results, we believe that semantic metrics are consistent with human evaluation. The model achieving the highest SCS and lowest HR is also the one most strongly preferred and rated as most adequate by participants. This suggests that semantic grounding is a key determinant of perceived description quality. Overall, with statistical significance, GPT-4o is the preferred and most adequate \gls{vlm}. Although the official size of GPT-4o is not publicly disclosed, it is estimated to be approximately three times larger than LLaVA and Qwen2-VL, which may explain these results. Thus, semantic metrics (SCS and HR) and VLM size serve as selection criteria for integrating future VLM models into our \gls{DRA} module, ensuring alignment with human judgment. Consequently, GPT-4o is selected as the VLM for \gls{DRA} in D-V2S. This matters because---though beyond this paper's scope---improved human comprehension of descriptions enables automotive engineers to perform targeted manual modifications (via natural language) for scenario variants absent from the original video. Note that this flexibility is precisely why we designed D-V2S as a two-stage pipeline with natural-language descriptions as the intermediate output. 

\begin{figure}
\centering
\setlength{\fboxsep}{4pt}
\fbox{
\begin{minipage}{0.9\columnwidth}
\begin{scriptsize}
1. The ego car is on the highway and the car in front of it suddenly decelerates.\\
2. The ego car that is on the left lane switches to the middle lane and another car on the right lane switches also to the middle lane.\\
3. Three pedestrians crossing in the middle of the road, one going from left to right and the two others from right to left.
\end{scriptsize}
\end{minipage}
}
\caption{Examples of the descriptions used as input to the \gls{SG}.}
\label{fig:descriptions-for-llm}

\end{figure}

\begin{table}
\centering
\footnotesize
\setlength{\tabcolsep}{5pt}
\caption{Description-to-scenario generation evaluation.}
\label{tab:scenario_eval}
\begin{tabular}{@{}lc@{}}
\toprule
\textbf{Method}
& \textbf{SP$\uparrow$}\\
\midrule
LCTGen           & 0.63 $\pm$ 0.35 \\
ChatScene        & 0.53 $\pm$ 0.31 \\
\gls{SG} (Ours)  & \textbf{0.93 $\pm$ 0.11}\\
\bottomrule
\end{tabular}

\end{table}

\begin{table}[t]
\centering
\footnotesize
\setlength{\tabcolsep}{2.5pt}
\caption{Scenario generation evaluation by the surveyed.}
\label{tab:human_eval_sg}

\begin{tabular}{lcccc}
\toprule
& \multicolumn{2}{c}{\textbf{Preference}} 
& \multicolumn{2}{c}{\textbf{Adequacy}} \\
\cmidrule(lr){2-3} \cmidrule(lr){4-5}
\textbf{Model} 
& Preference (\%) & 95\% CI (\%) 
& Mean $\pm$ SD & 95\% CI \\
\midrule
LCTGen    & 18.90 & 13.65--25.58 & 1.90 $\pm$ 1.51 & [1.67, 2.13] \\
ChatScene &  6.10 &  3.35--10.86 & 1.82 $\pm$ 1.15 & [1.64, 2.00] \\
Our       & \textbf{75.00} & 67.85--81.00 & \textbf{4.04 $\pm$ 1.29} & [3.84, 4.24] \\
\bottomrule
\end{tabular}

\vspace{0.25ex}
{\scriptsize
\textbf{\textit{Preference:}} \\
\textbf{\textit{Chi-square test:}} \\ 
$\chi^2(2, N=164)=132.16$, $p=2.0\times10^{-29}$, Cohen's $w=0.90$.\\
\textbf{\textit{z-test (p-value):}} \\
LCTGen vs ChatScene: $p=0.001$; \\ 
LCTGen vs Our: $p=7.4\times10^{-24}$; \\ 
ChatScene vs Our: $p=1.6\times10^{-36}$. \\

\textbf{\textit{Adequacy:}} \\
\textbf{\textit{One-way ANOVA:}} \\
$F(2,489)=148.6$, $p=3.8\times10^{-51}$, $\eta^2=0.378$.\\
\textbf{\textit{Post-hoc Tukey test (p-value) / Cohen's $d$:}} \\
LCTGen vs ChatScene: $p=0.85$, $d=0.06$; \\
LCTGen vs Our: $p<0.001$, $d=-1.52$; \\
ChatScene vs Our: $p<0.001$, $d=-1.82$. \\
}

\end{table}

\begin{table}
\centering
\caption{D-V2S Semantic Consistency.}
\label{tab:full_pipeline}
\begin{tabular}{lc}
\toprule
\textbf{Use case} & \textbf{E2E-SC $\uparrow$} \\
\midrule
Collision risk avoidance on highway & 0.86 $\pm$ 0.24 \\
Insertion on highway & 0.93 $\pm$ 0.24 \\
Pedestrian crossing in urban area & 0.80 $\pm$ 0.29 \\
Left Turn at urban intersection & 0.93 $\pm$ 0.18 \\
Pull back in on urban highway & 0.95 $\pm$ 0.16 \\
\midrule
Overall & 0.90 $\pm$ 0.21 \\
\bottomrule
\end{tabular}

\end{table}

\subsection{\glsdesc{SG} (\gls{SG})}

\subsubsection{SOTA} We consider LCTGen~\cite{LCTGen} and ChatScene~\cite{ChatScene2024}, which translate natural-language descriptions into scenarios. LCTGen uses GPT-4 to create structured vectors, retrieve road fragments, and render traffic via transformers. ChatScene parses descriptions into behavior/geometry/spawn clauses, matches SCENIC snippets from a database, and assembles executable scripts. For our \gls{SG}, we used GPT-4o \cite{GPT4} as the LLM. Since the different outputs are not directly comparable, we rendered each as bird's-eye views of dynamic participants' trajectories (Fig.~\ref{fig:qos_generation}).

\subsubsection{Descriptions} To evaluate \gls{SG} independently of \gls{DRA}, we compiled 23 scenario descriptions ({\eg},  Fig.~\ref{fig:descriptions-for-llm}): 15 with increasing complexity (single-vehicle to multi-vehicle/pedestrian) and 8 adapted from the Crash Report Dataset~\cite{NHTSA2016CIREN}.

\subsubsection{Experimental Setup and Evaluation Metrics} \Gls{SG} evaluation also uses semantic metrics and human judgment with statistical validation. We employ \emph{Semantic Preservation (SP)} (Table~\ref{tab:semantic_metrics}), which quantifies the proportion of relevant semantic structures faithfully reproduced during scenario synthesis---higher values indicating stronger fidelity between natural-language descriptions and generated scenarios. To compute SP, we define semantic elements $S_{\text{de}}$ (traffic participants, spatial relations, road topology, actions) from each description and extract generated elements $S_{\text{sc}}$ from the simulated scenario.

Additionally, we assessed perceived human alignment between descriptions and generated scenarios based on experienced drivers. We used a survey of 23 questions each time presented in random order and participants could stop anytime (no minimum answers). Scenario generation methods (Fig.~\ref{fig:qos_generation}) were anonymized and displayed in randomized positions to avoid biases. Each question required selecting the best-matching scenario (\emph{Preference}) and rating each scenario's adequacy on a 1--5 scale (\emph{Adequacy}). We collected 164 responses. 

\subsubsection{Results and Discussion}
Table~\ref{tab:scenario_eval} shows that \gls{SG} achieves a substantially higher SP (0.93) than LCTGen and ChatScene. \gls{SG} preserves substantially more semantic structure than competing methods. Table~\ref{tab:human_eval_sg} shows the preferences among surveyed: \gls{SG} led clearly at 75.0\%, followed by LCTGen (18.9\%) and ChatScene (6.1\%). The 95\% CI confirms these differences are meaningful---{\eg}, \gls{SG}'s CI (67.9--81.0\%) well exceeds the others. A chi-square test revealed highly significant deviation from equal preference ($\chi^2(2, N=164)=132.2$, $p<10^{-29}$, Cohen's $w=0.90$). Two-proportion $z$-tests confirmed \gls{SG} significantly outperformed LCTGen ($p<10^{-23}$) and ChatScene ($p<10^{-35}$). Table~\ref{tab:human_eval_sg} also reports adequacy scores: \gls{SG} received the highest ($4.04 \pm 1.29$), far exceeding LCTGen ($1.90 \pm 1.51$) and ChatScene ($1.82 \pm 1.15$). One-way ANOVA confirmed significant differences ($F(2,489)=148.6$, $p<10^{-51}$, $\eta^2=0.378$). Tukey tests showed that \gls{SG} rates significantly higher than competitors ($p<0.001$) and Cohen's $d$ confirmed large differences of \gls{SG} with them ($|d|>1.5$).

Hence, \gls{SG} outperforms LCTGen and ChatScene both semantically and according 
to human judgment. For instance, Fig. \ref{fig:qos_generation} shows that \gls{SG} successfully generated a realistic multi-vehicle interaction: vehicles are placed respecting the spatial constraints as stated in the natural-language description, and reproduced the described dynamics with fidelity.

\subsection{D-V2S (DRA + SG)}
\subsubsection{Experimental Setup and Evaluation Metrics} Direct comparisons with related methods were not feasible due to the absence of reproducible implementations or complete evaluation resources in existing works~\cite{dashcam2025,LEADE2026,Road2Code2025,Autoscenario2026}. To support reproducibility, we publicly release the complete D-V2S implementation\footnote{Code available at \href{https://alexandre-levy.github.io/DV2S.github.io/}{https://alexandre-levy.github.io/DV2S.github.io/}.}. We have considered the five use cases identified as especially relevant in the BERTHA Project \cite{BERTHA_D1_1}, with corresponding generated videos. The use cases are: (1) collision risk avoidance on highway, (2) insertion 
on highway, (3) pedestrian crossing in urban area, (4) left turn at urban 
intersection with oncoming cars, and (5) pull back in on urban highway.
For each use case we have 22 CARLA-generated videos, created by varying parameters such as vehicle speeds, spawn points for vehicles/pedestrians, and weather conditions. From these videos, D-V2S generates corresponding SCENIC scripts, which are executed in CARLA to produce output videos with a similar forward-facing horizontal FOV as the inputs. We annotated the relevant semantic elements in both input and output videos, enabling quantitative assessment of D-V2S. Specifically, we introduce the \emph{End-to-End Semantic Consistency (E2E-SC)} metric (Table~\ref{tab:semantic_metrics}), which directly compares the relevant semantic elements of input ($GT_{v}$) and corresponding output ($S_{o}$) videos.

\begin{table}[t]
\centering
\caption{Failure taxonomy of non-executable scripts.}
\label{tab:failure_taxonomy}
\begin{tabular}{l c}
\toprule
\textbf{Category} & \textbf{Description} \\
\midrule
Undefined symbol        & Code variable referenced but not declared \\
API hallucination       & Non-existent SCENIC attribute/method \\
Function misuse         & Missing required function parameter \\
Geometric infeasibility & Invalid spawn or spatial configuration \\
\bottomrule
\end{tabular}

\end{table}

\subsubsection{Results and Discussion}
Table~\ref{tab:full_pipeline} reports E2E-SC scores across evaluated use cases. For all of them, E2E-SC is higher than 80\%, with an overall value of 90\%. Results show that the observed CARLA-based simulations keep a strong semantic fidelity with the input videos. Moreover, this implies that the semantic information obtained by the \gls{DRA} module is effectively operationalized by the \gls{SG} module across essentially diverse scenarios. D-V2S took 19.08 seconds on average to convert each input video into a SCENIC script. On the other hand, we have also analyzed failure cases. The 94\% of generated SCENIC scripts compiled and executed successfully without modification. The remaining 6\% had minor syntactic inconsistencies or missing parameters (see Table~\ref{tab:failure_taxonomy}), not structural semantic errors. Failures stemmed from code formatting rather than scene misinterpretation.

\begin{figure}
    \centering
    \includegraphics[width=0.475\textwidth]{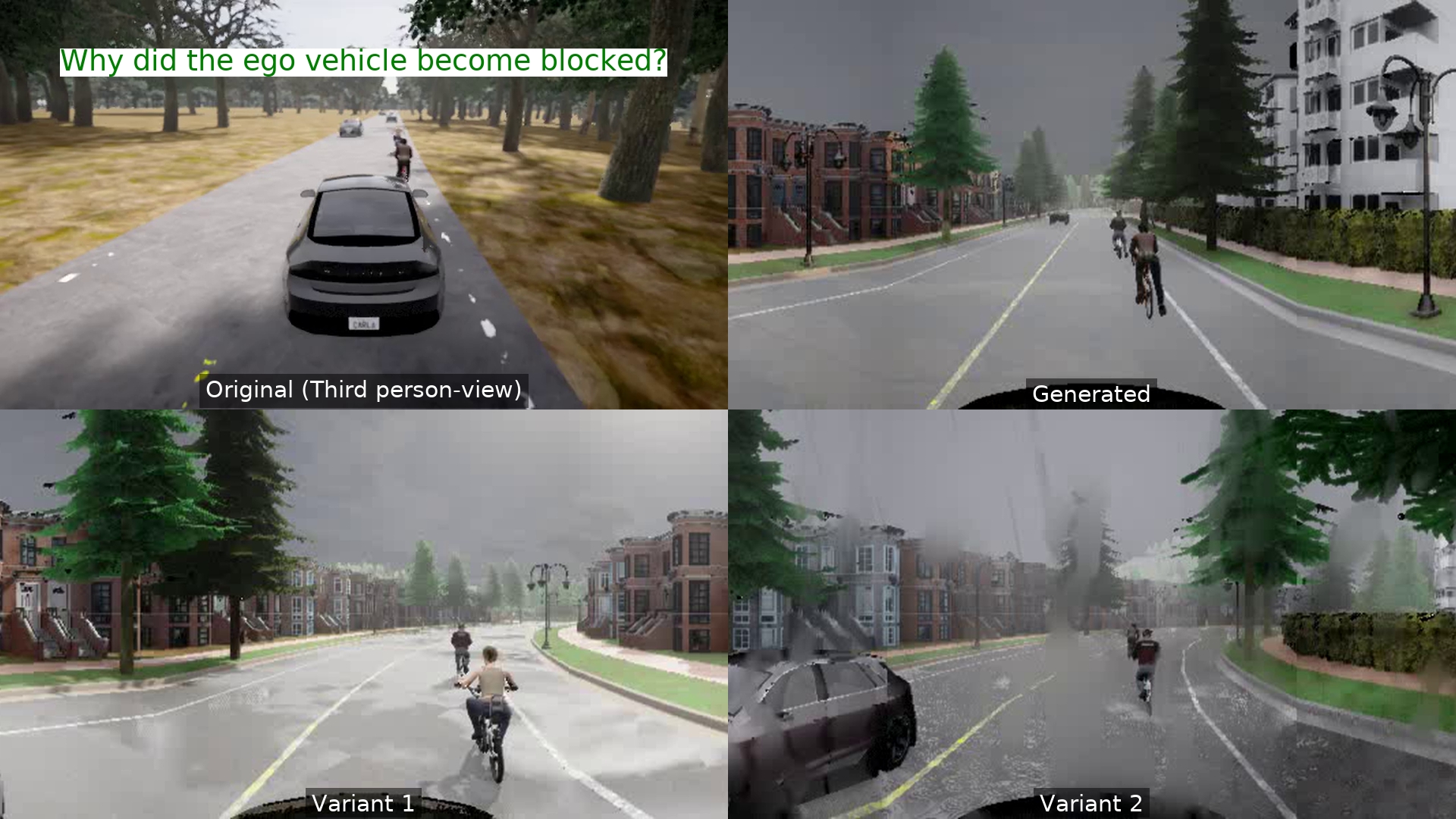}
    \caption{D-V2S transforms a third-person perspective video from CARLA, captured during goal-oriented autonomous driving (top-left), into a corresponding SCENIC script. The overlaid green text is the scenario-specific user question (green prompt part, Fig.~\ref{fig:prompt}). The top-right shows a CARLA-generated video obtained by executing this D-V2S-generated SCENIC script using an onboard camera perspective. Bottom frames depict variants of the original scenario, generated by modifying SCENIC script parameter values such as weather conditions and specific traffic actors within meaningful intervals (which can be randomly done or as an user requirement).}
    \label{fig:quali_synth}

\end{figure}

\begin{figure}
    \centering
    \includegraphics[width=0.475\textwidth]{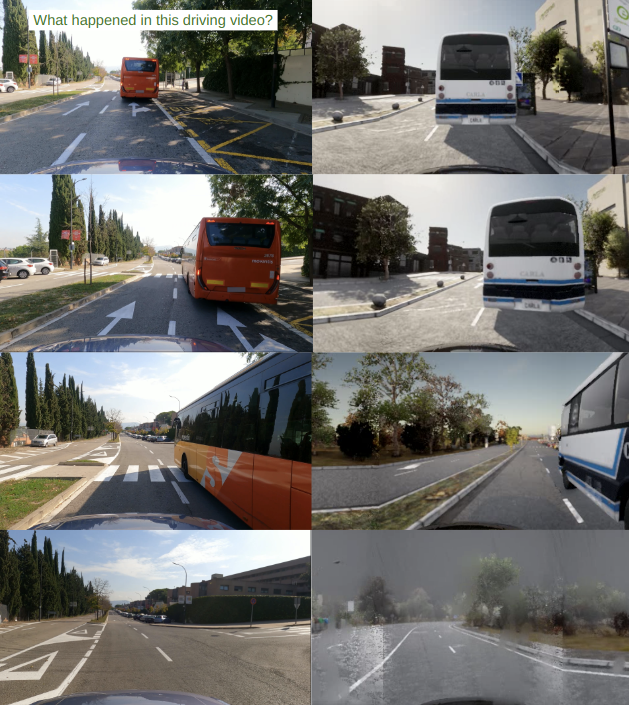}
    \caption{Left column: GoPro frames (windshield-mounted). D-V2S generates a SCENIC script from this video and prompt (green, top image). Right column: executing the script in CARLA reproduces a semantically similar scenario, using three forward-facing cameras to match GoPro's fish-eye field-of-view. Rows 1-2 show the same simulation; rows 3-4 show variants with different illumination/weather, which is straightforward to do once the SCENIC script exists.}
    \label{fig:quali_real}

\end{figure}

Qualitative results are shown in figures \ref{fig:quali_synth} and \ref{fig:quali_real}, the latter demonstrating transferability 
to realistic conditions using windshield-mounted GoPro footage. The 
third-person camera perspective is used only for interpretability, 
not as a framework requirement.

Together, our quantitative and qualitative results demonstrate that D-V2S represents a significant advance in automatically generating semantically faithful simulation scenarios from videos, accurately reproducing the relevant content of the original footage while enabling variants via natural-language descriptions (output of \gls{DRA}) or SCENIC script parameter value modifications (output of \gls{SG}).

\section{Conclusions}
\label{sc:con}

This paper presented D-V2S, a novel two-stage framework that automatically transforms driving videos---real or simulated---into executable SCENIC scenarios. By leveraging off-the-shelf VLMs and LLMs with carefully designed prompts---no retraining required---D-V2S preserves an average of 90\% of relevant semantic elements while enabling editable outputs. Ablation studies confirm GPT-4o's superiority in DRA (semantic accuracy and human preference) and D-V2S's SG outperforming LCTGen and ChatScene (75\% preference rate). These results demonstrate D-V2S's effectiveness for scalable generation of semantically faithful scenarios---preserving key events and interactions among participants---advancing interpretable testing for autonomous vehicles. Future work will integrate D-V2S in \gls{ai} driving development loops and address current failure cases.

\section*{Acknowledgments}
Funded by the European Union under Grant Agreement 101076360 (BERTHA). Views and opinions expressed are however those of the author(s) only and do not necessarily reflect those of the European Union or the European Climate, Infrastructure and Environment Executive Agency (CINEA). Neither the European Union nor the granting authority can be held responsible for them. BERTHA was particularly relevant for the definition of the driving scenarios considered in this work. A.M. L\'opez acknowledges financial support for his general research activities from ICREA under the ICREA Academia Program. All authors also acknowledge the support of the Generalitat de Catalunya through the CERCA Program and its ACCI\'O Agency for CVC’s general activities. Authors acknowledge fruitful discussions on: a) foundation models carried out within ELLIOT (Funded by the European Union under Grant Agreement 101214398) and b) explainability for autonomous driving within the HAMILTON project (PID2024-157936NBI00), funded by MICIU/AEI/10.13039/501100011033 and by ERDF, EU.

\hypersetup{urlcolor=black}
\balance
\bibliography{arxiv}

\end{document}